\documentclass[12pt]{article}

\usepackage{geometry}
\usepackage{booktabs}
\usepackage{longtable,lscape}
\usepackage{graphicx}
\usepackage{multicol}
\usepackage{setspace}
\usepackage[UKenglish]{babel}
\usepackage{natbib}
\usepackage{tabularx}
\usepackage{multirow}
\usepackage{hyperref}

\def\b#1{\mbox{\boldmath $#1$}}    
\def\bl#1{\mbox{\scriptsize \boldmath {$#1$}}} 
\def\cg#1{\mbox{${\cal #1}$}}      

\renewcommand{\th}{\theta}

\newcommand{\la}{\lambda}

\def\b#1{\mbox{\boldmath $#1$}}

\begin{document}
\begin{center}

{\Large \bfseries Item selection by Latent Class-based methods} \vspace{10 mm}

{\large Francesco Bartolucci, Giorgio E. Montanari, Silvia Pandolfi
}

\vspace{10 mm}

{\large \today}

\vspace{10 mm}

\end{center}
\begin{abstract}
\begin{singlespace}

The evaluation of nursing homes is usually based on the administration of questionnaires made of a large number of polytomous items. In such a context, the Latent Class (LC) model represents a useful tool for clustering subjects in homogenous groups corresponding to different degrees of impairment of the health conditions. 
It is known that the performance of model-based clustering and the accuracy of the choice of the number of latent classes may be affected by the presence of irrelevant or noise variables. In this paper, we show the application of an item selection algorithm to real data collected within a project, named ULISSE, on the quality-of-life of elderly patients hosted in italian nursing homes. This algorithm, which is closely related to that proposed by Dean and Raftery in 2010,   
 is aimed at finding the subset of items which provides the best clustering according to the Bayesian Information Criterion. At the same time, it allows us to select the optimal number of latent classes. 
Given the complexity of the ULISSE study, we  perform a validation 
of the results by means of a sensitivity analysis to different specifications
of the initial subset of items and of a resampling procedure. 

\vspace*{0.25cm}
\end{singlespace}

\end{abstract}
\newpage

\section{Introduction}\label{sec:intro}

The evaluation of appropriateness of long-term care facilities is
assuming a role of increasing relevance due to the rapid growth of
demand for long-term care services for elderly people. The main
cause is represented by the rapid aging of the population and also
by the changes in the family structure and in the socio-economic
context. Furthermore, the debate on population
aging focuses on the effects that such a phenomenon has on the
welfare and on the health care system in various countries
\citep{gal:prof:07,Brey:et:al:2010}. In this regard, health care
quality measurement and performance evaluation of nursing homes
represent a challenging issue to assure the quality of services and
to allocate resources efficiently. 


Issues related to population aging are particularly relevant in
Italy, which is one of the European countries with the highest
proportion of elderly people, where this proportion is expected
to increase over the next decades \citep{kohler_et_al2002}. 
In this country, the ULISSE
project \citep{lattanzio2010} has been carried out to obtain
relevant data for health care planning. The purpose of the project is to document
the change in elderly patients' health status and the ability of the
health care system to satisfy their needs. The dataset obtained from
this project was collected by the administration of a questionnaire
to patients hosted in a sample of Italian nursing homes. The
questionnaire is made of a large number of polytomous items about different aspects of the
quality-of-life and health status of these patients. In such a context, a model-based clustering 
procedure may be applied to evaluate the performance of nursing homes. In particular,
patients may be clustered in homogeneous groups according to their health conditions in order to
describe the case-mix of the nursing homes. 
The resulting clustering may have important applications in the context of the nursing
home evaluation, and therefore affecting the system of financial support, when the clusters correspond to different
degrees of impairment of the patients' health conditions. 
However, the performance of model-based clustering procedure may be degraded by the presence of irrelevant items. Moreover, the 
administration of a questionnaire made of a large number of items may be lengthy and expensive. Due to tiring effects,
using several items may also induce the respondent to
provide inaccurate responses. This is particularly relevant when the
questionnaire is periodically administered. Therefore, methods which
allow us to select the smallest subset of items useful for
clustering are of interest. These methods may lead to a reduction of
the costs of the data collection process, and a better quality of
the collected data. Moreover, reducing the dimension of the dataset
implies that it may be more easily analyzed by complex statistical
models.

Motivated by availability of the ULISSE dataset, in this paper we
adopt the item selection algorithm proposed by \cite{dean:raftery:2010}, 
which may be applied when a
large number of items is included in a questionnaire. 
The adopted algorithm is based on the latent class (LC) model
\citep{laza:50,laza:hen:68,good:74} which represents an important tool of
analysis of data collected by questionnaires made of polytomous
items. As it is well known, the model relies on a discrete latent
variable, which defines a certain number of latent classes, and it
assumes the independence of the responses to the items given this
variable. Therefore, the use of the LC model is justified when the
items measure one or more latent traits, such as the quality-of-life
or the tendency toward a certain behavior. In geriatrics, the LC
model is used to measure mobility disability \citep{Band:et:al:97},
to study behavioral syndromes in Alzheimer' patients
\citep{moran:et:al:04}, and to test the validity of certain physical
frailty measures \citep{Band:et:al:06}. Moreover,
\cite{lafor:et:al:09} uses the LC analysis to model the
heterogeneity in elderly individuals' health status.

It is important to recall that the LC model produces a {\em
model-based clustering} \citep{fraley:raftery:2002}, with clusters
corresponding to the latent classes. Once the model is fitted, a
subject is assigned to the latent class corresponding to the highest
posterior probability, that is, the conditional probability of the
latent class given the observed data. 
It is also worth noting that latent variable models for ordinal
variables are also present in the literature, such as the graded
response model \citep{sam:69,sam:96}. In this model, the response
probability is expressed as a function of one or more latent
variables through a specific link function (e.g., the cumulative
logit link); see \cite{bacci:et:al:14} and the reference therein.
Given the ordinal nature of the items composing the questionnaire
considered in our application, this model could make sense in the
present context. However, we prefer to avoid such a parametrization and
to rely on a standard LC model. The main advantages of this choice are the
simplicity of the resulting approach and the reduced number of parametric
assumptions.

The item selection algorithm we adopt aims at finding the set of
items which provides the best value of the Bayesian Information
Criterion (BIC) index \citep{schw:78,kass:raftery:1995}. 
As motivated by \cite{dean:raftery:2010}, this leads to selecting the subset of items which are indeed useful for clustering. It is
worth noting that, at the same time, this method allows us to choose
the number of latent classes. As usual in the context of LC models, the parameters estimation is 
performed by the
Expectation-Maximization algorithm
\citep[EM;][]{good:74,demp:et:al:77}.

More in detail, the implementation of the 
algorithm is based on a stepwise scheme that, starting from an initial set of items, 
at each iteration performs both inclusion and exclusion steps till a certain optimal criterion is satisfied.
We also extend the version proposed in \cite{dean:raftery:2010} by including
an additional step aimed at initializing, with a large number
of starting values, the estimation algorithm, so as to face the problem of
local maxima of the model log-likelihood, without being computationally expensive. This problem is particularly evident
in applications involving a very large number of statistical units and response variables, as
in the ULISSE project.  In such an application, we also assess the performance of the 
item selection algorithm by implementing a sensitivity analysis of the final results with respect to different specifications of the initial subset of items and by validating the solution on the basis of resampling procedures.

The remainder of this paper is structured as follows. The ULISSE
dataset is described in the following section. In
Section~\ref{sec:LC} we briefly illustrate the LC model on which the
item selection algorithm is based together with maximum
likelihood estimation of the model, via the EM algorithm.
In Section~\ref{sec:item} we illustrate the item selection algorithm
based on the implementation proposed by \cite{dean:raftery:2010}. In
Section~\ref{sec:Ulisse} we present the results of this
approach applied to the ULISSE dataset, whereas 
Section~\ref{sec:concl} provides a final discussion.

The approach proposed in this paper has been implemented in
a series of {\sc R} functions which rely on a Fortran code to make
the execution faster. These functions are available to the reader
upon request.
\section{The ULISSE dataset}\label{sec:dataset}
The ULISSE project (``Un Link Informatico sui Servizi Sanitari
Esistenti per l'Anziano'' - ``A Computerized Network on Health Care
Services for Older People'') is aimed at studying the health status
of elderly patients who currently receive health care assistance in
Italy \citep{lattanzio2010}. The main purpose of the study is to
improve the knowledge of the characteristics and the quality of
health care services provided to elderly in Italy. The
project was carried out by a Research Group established by the
Italian Ministry of Health. The study
considers three different levels of health care assistance: that
provided by acute care facilities, that provided by nursing homes,
and that provided at home. Overall, 23 acute geriatric or internal
medicine hospital units, 31 nursing homes, and 11 home care services
have been involved in the project. In the analysis here presented, we
consider only data collected in the nursing homes.

The ULISSE project is based on a longitudinal survey; in the nursing
homes recruited by the project, all residents, or a maximum of 100
randomly selected residents for bigger nursing homes, were evaluated
at admission and then re-evaluated at 6 and 12 months after the
admission. Only long stay residents (i.e., permanently admitted to
the nursing home) aged at least 65 years were included in the study.
For the analysis here presented, we consider only the
first wave of interviews, which covers 1739 patients.

The detailed patients information were collected using the
classification system VAOR (``Valutazione dell'Anziano Ospite di
Residenza'') that represents the Italian version of the interRAI
Minimum Data Set (MDS) for nursing home care
\citep{morris_et_al1991,haw:et:al:97}. The questionnaire, which was filled up by the nursing
assistants, is made of 75 items covering
several aspects of the health status of the elderly patients.
The items are polytomous, with categories generally ordered
according to increasing difficulty levels in accomplishing a certain
task or severeness of a specific aspect of the health conditions. The complete list of
items, with the corresponding number of response categories, is
reported in Appendix.

The 75 items are grouped into eight different sections
(indexed by $d$) of the questionnaire, concerning:
\begin{enumerate}
\item  Cognitive Conditions (CC);
\item Auditory and View Fields (AVF);
\item Humor and Behavioral Disorders (HBD);
\item Activities of Daily Living (ADL);
\item Incontinence (I);
\item Nutritional Field (NF);
\item Dental Disorders (DD);
\item Skin Conditions (SC).
\end{enumerate}

Table~\ref{tab:mis} shows the average of
the percentage of missing values computed with respect to the items
composing each section of the questionnaire. We observe that the percentage of missing responses is considerably
different between sections, going from 0.54\%
for section AVF to 9\% for section ADL.

\begin{table}[ht]\centering\vspace*{0.25cm}
\small{
\begin{tabular}{clr}
\toprule $d$ & \multicolumn1c{section} & \multicolumn1c{\% missing} \\
\midrule
1 & CC & 0.78 \\
2 & AVF & 0.54 \\
3 & HBD & 1.96 \\
4 & ADL & 9.00 \\
5 & I & 1.61 \\
6 & NF & 5.58 \\
7 & DD & 2.18\\
8 & SC & 6.48 \\
\bottomrule
\end{tabular}
\caption{\em Average percentage of missing values for each section
of the questionnaire.}\label{tab:mis}}\vspace*{0.25cm}
\end{table}%

\section{The Latent Class model}\label{sec:LC} 

Let $J$ denote the number of items in the questionnaire of interest
and, for a sample of $n$ respondents, let $Y_{ij}$ denote the
response variable for subject $i$ and item $j$, with $i=1,\ldots,n$
and $j=1,\ldots,J$. Let $l_j$ denote the number of response categories of
item $j$, labeled from 0 to $l_j$-1. Some of the responses may not
be observed for different reasons.  We rely on the standard assumption of {\em missing at random} 
\citep[MAR;][]{rubin:76,litl:rubin:1987} to deal with these missing responses, as described in the following.
Finally, let $\b
Y_i=(Y_{i1},\ldots,Y_{iJ})$ be the vector of all response variables for subject $i$.

In order to explain the dependence structure between the response
variables, the LC model assumes the existence of a discrete latent
variable $U_i$ which has, {\em a priori}, the same distribution for
every subject $i$. This distribution is based on $k$ support points,
labeled from 1 to $k$. Each support point corresponds to a latent
class in the population and has a specific weight (or {\em a priori}
probability); these weights are denoted by $\pi_1,\ldots,\pi_k$.
Moreover, the conditional probability that individual $i$ in class
$u$ provides response $y$ to item $j$ is
\[
\lambda_{j|u}(y)=p(Y_{ij} = y|U_i = u),\quad
j=1,\ldots,J,\:u=1,\ldots,k,\:y = 0,\ldots,l_j-1.
\]
Overall, the number of
non-redundant parameters of the model is
$$
g_k = (k-1) + k \sum_j (l_j-1).
$$
The basic assumption underlying the LC model is that of {\em local independence}. This assumption is formulated by requiring that,
for $i=1,\ldots,n$, the variables
$Y_{i1},\ldots,Y_{iJ}$ are conditionally independent given the latent
variable $U_i$. 
This assumption implies that
\[
p(\b y_i|u) = P(\b Y_i = \b y_i| U_i = u) = \prod_{j=1}^J \lambda_{j|u}(y_{ij}),
\]
with $\b y_i = (y_{i1},\ldots,y_{iJ})$. Moreover,
the {\em manifest probability} of the
response pattern $\b y_i$ for subject $i$ is denoted by
\[
p(\b y_i)  =  \sum_{u=1}^k p(\b
y_i|u)\pi_u.
\]
Another quantity of interest is the posterior probability that a
subject with observed response configuration $\b y_i$
belongs to latent class $u$. Using standard rules, this probability
is equal to
\begin{equation}\label{eq:post_prob}
p(u|\b y_i)=\frac{p(\b
y_i|u)\,\pi_u}{p(\b y_i)},\quad u=1,\ldots,k.
\end{equation}
These posterior probabilities are used to allocate subjects in the
different latent classes, as will be clarified in the sequel.

As defined above, in the presence of missing responses to the items, we rely on the MAR hypothesis.
This assumption states that the probability of the observed
missingness pattern, given the observed and the unobserved data,
does not depend on the unobserved data \citep[see, among
others,][]{lu:copas:04}. Therefore, provided that the model for the
missing data mechanism is separated from that of the LC model, the
missing responses are {\em ignorable} for likelihood-based
inference. The corresponding LC model may be formulated by introducing the 
missing data indicator $M_{ij}$ 
which is equal to 1 when subject i does not respond to item $j$ and to 0 otherwise.
Then, for a given subject $i$, we have $\b M_i = (M_{i1}, \ldots, M_{iJ})$. The corresponding
response pattern is given by $(\b m_i,\b y_{i,obs})$ in which $\b m_i$ is a configuration of $\b M_i$ and the subvector $\b y_{i,obs}$
contains the observed components of $\b Y_i$.

The MAR assumption implies that the parameters may be estimated on the
basis of the log-likelihood of the vectors of observed responses $\b
y_{i,obs}$ only, without worrying about the model for missingness \citep[see,among others,][]{har:sch:09}.
In particular, the
conditional probability of the observed response configuration,
given the latent class, is simply
$$p(\b y_{i,obs}|u) = \int_{\bl y_{i,mis}} \!p(\b y_i|u)\; \mathrm{d}\,\b y_{i,mis} =\prod_{j:m_{ij}=0}
\lambda_{j|u}(y_{ij}),$$ and then
\begin{equation}\label{eq:manif_MAR} p(\b y_{i,obs}) = \sum_u p(\b
y_{i,obs}|u)\pi_u.
\end{equation}
Note that under this assumption the number of parameters to be estimated remains the same 
as in the standard LC model.

\subsection{Maximum likelihood estimation}\label{sec:EM}
Given the assumption of independence between the sample units, the general formulation of the log-likelihood
function of the proposed model is
\[
\ell(\b\th)=\sum_i \log p(\b y_i). 
\]
In the presence of missing values considered as MAR, the above
expression becomes
\[\ell(\b\th) = \sum_i
\log p(\b y_{i,obs}),\] 
where the manifest probability
$p(\b y_{i,obs})$ is computed as in (\ref{eq:manif_MAR}). 
In the above expressions, $\b\theta$ is a short-hand notation for all model parameters.
In order to estimate these parameters, we maximize $\ell(\b\th)$ by
the EM algorithm \citep{demp:et:al:77}. 

\subsection{Expectation-Maximization
algorithm}\label{sec:EMalgorithm}

The EM algorithm is based on the {\em complete-data likelihood} that
we could compute if we knew the value of the latent variable $U_i$
for every unit $i$ in the sample. This is equivalent to the
knowledge of the latent class to which every subject belongs. As usual, we
represent such an information by the set of dummy variables $z_{iu}$,
$i=1,\ldots,n$, $u=1,\ldots,k$, where $z_{iu}$ is equal to 1 if
respondent $i$ belongs to latent class $u$ and to 0 otherwise. Then,
under the MAR assumption,
we can write the complete-data log-likelihood as
\begin{eqnarray}
\ell^*(\b\theta)&=&\sum_i\sum_u z_{iu}\log[p(\b y_{i,obs}|u)\pi_u] \nonumber\\
&=&\sum_i\sum_u z_{iu}\sum_{j:m_{ij}=0}\log \la_{j|u}(y_{ij})+ \sum_u
z_{+u}\log \pi_u, \label{eq:comp_lk}
\end{eqnarray}
where $z_{+u}=\sum_iz_{iu}$ is the number of subjects in latent
class $u$. We have an explicit solution for the maximum of
$\ell^*(\b\theta)$ with respect to the model parameters, which is
\begin{eqnarray}
\tilde{\pi}_u & = & \frac{z_{+u}}{n},
\quad u=1,\ldots,k, \label{eq:Mstep1}\\
\tilde{\lambda}_{j|u}(y) & = & \frac{\sum_i
I(y_{ij}=y)(1-m_{ij})z_{iu}}{\sum_i z_{iu}(1-m_{ij})},\quad
j=1,\ldots,J,\,u=1,\ldots,k,\,y=0,\ldots,l_j-1,
\label{eq:Mstep2}\end{eqnarray} where $I(\cdot)$ is the indicator
function equal to 1 if its argument is true and to 0 otherwise.

In order to maximize the model log-likelihood, the EM algorithm
alternates the following two steps until convergence, starting from
an initial guess of the model parameters in $\b\th$:
\begin{itemize}
\item {\bf  E-step}: compute the conditional expected value of the complete-data
log-likelihood $\ell^*(\b\th)$ given the observed data and the current value of the
parameters;
\item {\bf M-step}: update the model parameters by maximizing the expected
value obtained at the E-step.
\end{itemize}

Both steps are simple to implement. In practice, the E-step consists
of obtaining the posterior expected value of every dummy variable
$z_{iu}$, that is,
\[
\hat{z}_{iu}=p(u|\b y_i), \quad i=1,\ldots,n,\:u=1,\ldots,k,
\]
which may be computed according to (\ref{eq:post_prob}). At the
M-step we maximize the expected value of the complete-data
log-likelihood, which is obtained by substituting every dummy
variable $z_{iu}$ in (\ref{eq:comp_lk}) with $\hat{z}_{iu}$, and in
this way we update the parameter vector $\b\th$.  For this
maximization we use formulae (\ref{eq:Mstep1}) and
(\ref{eq:Mstep2}), with $\hat{z}_{iu}$ instead of $z_{iu}$.

As mentioned in Section \ref{sec:LC}, the posterior probabilities
$\hat{z}_{iu}$ can be used for clustering, that is, to allocate
subjects in the $k$ latent classes. In particular, on the basis of
the output of the EM algorithm, we assign subject $i$ to latent
class $u$ when $\hat{z}_{iu}=\hat{z}_i^*$, where $\hat{z}_i^*$ is
the maximum of $\hat{z}_{i1},\ldots,\hat{z}_{ik}$. For this reason,
\cite{mag:ver:01} and \cite{ver:mag:02} refer to this kind of model
as an {\em LC cluster model}.

 

%
\subsubsection{Initialization of the algorithm}\label{sec:init}

A typical problem of the latent variable and finite mixture models
is the multimodality of the likelihood. Obviously, in the presence
of multiple local maxima, the EM algorithm converges to one of them,
which is not ensured to be the global maximum. In this case, it is
advisable to use 
a random
initialization strategy which consists of repeatedly
initializing the algorithm from a large number of randomly chosen starting values for
the parameters.  When more starting values are used, the final
estimate is the one corresponding to the largest likelihood value
that has been found at convergence of the EM algorithm
\citep{bier:et:al:03,karlis:xeka:03}. This solution is not
guaranteed to correspond to the global maximum; however, it is
rather obvious that the chance of reaching the global maximum
increases with the number of the starting values that are tried. The problem
of likelihood multimodality is particularly severe for the LC model
used in our analysis, due to the large number of items and to the
fact that these items generally have more than two response
categories.

In our approach we adopt a random initialization which is based on
drawing each latent class weight $\pi_u$ from a uniform distribution
between 0 and 1 and then normalizing these random draws so that they
sum to 1. In a similar way we randomly choose the conditional response probabilities
$\la_{j|u}(y)$, $y=0,\ldots,l_j-1$. Moreover, we use a number of random
initializations which increases with the number of latent classes,
because the latter affects the number of parameters and then the
expected number of local maxima.

\section{Item selection procedure} \label{sec:item}
In this section we describe the procedure for item selection based on the
method proposed by \cite{dean:raftery:2010}. This method follows a
stepwise scheme that, starting from an initial set of items,
performs a series of inclusion and exclusion steps until
a suitable stopping rule is satisfied, also allowing the selection of the proper
number of latent classes. 

\subsection{Inclusion-exclusion algorithm with random check}\label{sec:inc_exc}

The inclusion-exclusion algorithm for item
selection is based on assessing the importance of a certain item by
comparing two LC models. In the first model, the item is assumed to
provide additional information about clustering allocation, beyond
that contained in the already selected items; in the second model,
this item does not provide additional information useful for
clustering and then it is independent of the latent variable
defining the latent classes. The two models are compared via BIC
index \citep{schw:78}, which is seen as an approximation of the Bayes Factor
\citep{kass:raftery:1995}.

In more detail, let $\cg A^{(0)}$ denote the initial set of items and let 
$k^{(0)}$ the  corresponding number of latent classes.
At the $h$th iteration, the item selection algorithm performs an
inclusion and an exclusion step, so as to obtain $\cg
A^{(h)}$ and $k^{(h)}$,  as follows:
\begin{itemize}
\item {\em Inclusion step}: each item $j$ in $\bar{\cg A}^{(h-1)}$, the complement of $\cg A^{(h-1)}$ with
respect to the full set of items, is singly proposed for inclusion
in $\cg A^{(h)}$. The item to be included is selected on the basis
of the difference between BIC of the LC model for the items in
$\cg A^{(h-1)} \cup j$ (optimized over the number of classes from 2 to $k_{\max}$, with $k_{\max}$ {\em a priori} fixed) and BIC of the LC model in which item $j$ is
assumed to be independent of the latent class. This difference is as follow
\[
BIC_{diff}(\cg A^{(h-1)},j)=\min_{2\leq k \leq k_{\max}}BIC_k(\cg A^{(h-1)} \cup
j)-[BIC_1(j) + BIC_{k^{(h-1)}}(\cg A^{(h-1)})],
\]
where
\begin{equation}\label{eq:BIC} BIC_k(\cg A) = -2
\;\hat{\ell}_k(\cg A) + g_k(\cg A)\log(n),\end{equation} with
$\hat{\ell}_k(\cg A)$ being the maximum of the log-likelihood of the
LC model applied to the data referred to the items in $\cg A$, and
$g_k(\cg A)$ being the corresponding number of free parameters.
Obviously, $BIC_1$ is the BIC index of the single class LC model,
which corresponds to the model of independence. Note that each
step of the algorithm also selects the number of classes $k$, since $BIC_k$ is minimized over $k$ from 2 to $k_{\max}$.
The item included is the one with the smallest negative $BIC_{diff}(\cg A^{(h-1)},j)$, and $\cg A^{(h)}=\cg A^{(h-1)} \cup j$.
If no item 
yields a negative $BIC_{diff}$, then we set $\cg A^{(h)}= \cg A^{(h-1)}$.
\item {\em Exclusion step}: each item $j$ in $\cg A^{(h)}$ is singly proposed for exclusion.
The item to be removed from $\cg A^{(h)}$ is selected on the basis
of the same criterion as above, that is,
\[
BIC_{diff}(\cg A^{(h)}\setminus j,j)=BIC_{k^{(h)}}(\cg A^{(h)})-[BIC_1(j) + \min_{2\leq k \leq k_{\max}}BIC_k(\cg A^{(h)}\setminus j)].
\]
The item with the highest positive value of $BIC_{diff}(\cg A^{(h)}\setminus j,j)$ is removed from $\cg A^{(h)}$ and $k^{(h)}$ is updated. If no item is found
with a positive $BIC_{diff}(\cg A^{(h)}\setminus j,j)$, $\cg A^{(h)}$ is left unchanged.
\end{itemize}
The algorithm ends when no item is added to $\cg A^{(h)}$ and no item is removed from
$\cg A^{(h)}$. It has to be clear that different LC models are estimated at each
step of this algorithm. These models are different in the set of
items and in the number of latent classes. Obviously, an initialization of the
EM algorithm is required for each of these models. In particular,
\cite{dean:raftery:2010} propose to use the parameter estimates, available at
the end of the previous step of the item selection algorithm, to
obtain these starting values. In more detail, they use the estimated
posterior probabilities $\hat{z}_{iu}$ as reasonable starting
values for models involving the updated dataset, with one more or
one less item. When a different number of latent classes is
considered, 
the new latent classes are
obtained by collapsing  two or more closest classes, in terms of
Euclidean distance between the corresponding conditional response
probabilities, or by splitting one or more classes into new classes. 

As the above initialization does not prevent the problem of the likelihood multimodality that is particularly
severe for the LC model, we propose to include an additional step, after each inclusion and
exclusion step
aimed at performing a check based on several random initializations of the EM algorithm. 
In particular, as outlined in Section \ref{sec:init},
we initialize the EM algorithm by a large number of random starting values, proportional to 
the current number of latent states, and we take the estimates corresponding
to the highest log-likelihood at convergence of the algorithm. 
This random check, which is performed once an item has been included or removed 
and the corresponding number of latent classes has been selected, allows us to assess the convergence
to the global maximum of the model likelihood, without being too computationally expensive. 


The item selection algorithm at issue also requires to properly
choose the initial set $\cg A^{(0)}$ of items. For this aim,
\cite{dean:raftery:2010} propose to estimate an LC model with at
least 2 classes for all the items and to order these items in terms
of variability of the corresponding estimated response probabilities
across classes. Then, they choose the smallest set of items that,
among the items having the highest variability, allows at least a
2-class model to be identified. In the application which follows, we also perform 
a sensitivity analysis of the final solution to different initial set of items. 

It is worth noting that $BIC_{diff}(\cg A,j)$ corresponds to the
difference between the BIC index for an LC model based on the
assumption that the items in $\cg A \cup j$ depend on the latent
variable (and the items in $\bar{\cg A}\setminus j$ do not depend on
the latent variable) and that for the corresponding model in which only
the items in $\cg A$ depend on the latent variable (and the items in
$\bar{\cg A}$ do not depend on the latent variable). Consequently, the
item selection algorithm above is aimed at finding the set of items
and the number of classes which minimize the index:
\begin{equation}\label{eq:BICtot}
BIC_{tot,k}(\cg A) = BIC_k(\cg A)+BIC_1(\bar{\cg A}).
\end{equation}
Through the last index it is then possible to compare the solutions obtained by the item selection
algorithm, even with a different number of latent classes or a
different number of items included in the set $\cg A$.

\section{Application to the ULISSE dataset}\label{sec:Ulisse}
In this section, we illustrate the results obtained from the
application of the item
selection algorithm to the ULISSE dataset described in
Section~\ref{sec:dataset}.  

\subsection{Item selection}\label{sec:item_sel}

First of all, in order to apply the inclusion-exclusion algorithm,
we fit the LC model described in Section~\ref{sec:LC},
considering the full set of 75 items, denoted by $\cg J$, for a
number of latent classes from 2 to $k_{\max}$, with $k_{\max}=10$.
For each $k$, we initialize the EM algorithm by means of $100\times(k-1)$ random 
starting values. The number of latent classes corresponding to the minimum of
$BIC_k(\cg J)$ defined in (\ref{eq:BIC}) is $k=8$.

In order to select the initial set of clustering items, for each category
of each item, we then calculate the variance of its estimated conditional 
probability across classes. For each item, we finally add up these variances
and we order the items according to this sum. This because items with high 
values of this sum have high between-class variability, and therefore may be 
more useful for clustering.
We consider
different sizes of the initial set $\cg A^{(0)}$, equal to
$3,10,20,30,75$, to perform the inclusion-exclusion algorithm. Using
different choices allows us to study the sensitivity of the final
solution with respect to different starting sets. For each initial set of items, we
select the initial number of latent classes $k^{(0)}$, always on the basis of 
$BIC_k(\cg A^{(0)})$, for $k=2,\ldots,k_{\max}$, and we apply the item selection procedure.

Table~\ref{tab:DeR} reports the results obtained, in
terms of number of items, number of classes, and corresponding
$BIC_k(\hat{\cg A})$ and $BIC_{tot,k} (\hat{\cg A})$ index, defined in (\ref{eq:BICtot}), of the final model. 
The output of the inclusion-exclusion algorithm
shows a moderate dependence on the choice of the initial set of
clustering items. Moreover, the random check allows us to increase the chance of reaching a global 
maximum of the model log-likelihood, with always better results in terms of $BIC_{tot,k}(\hat{\cg A})$ with 
respect to the algorithm without the random check.
From the results, we also note that the selected items are mostly equivalent, apart from four items, 
and the number of classes varies from $\hat{k}=8$ and $\hat{k}=10$, leading to different values of the $BIC_{tot,k}(\hat{\cg A})$
index. The best result, in
terms of $BIC_{tot,k}(\hat{\cg A})$, is obtained with an initial set of 30 items,
 that leads to selecting 50 items and $\hat{k}=9$ latent
classes.

\begin{table}[h!]\centering\vspace*{0.25cm}
\small{
\begin{tabular}{rrccrrr}
\toprule
\multicolumn{1}{c}{size of $\cg A^{(0)}$} &\multicolumn{1}{c}{ $k^{(0)}$} &{\em random check} & \multicolumn{1}{c}{\# items} & \multicolumn{1}{c}{ $\hat{k}$} & \multicolumn{1}{c}{$BIC_{k}(\hat{\cg A})$} & \multicolumn{1}{c}{$BIC_{tot,k}(\hat{\cg A})$}\\   
\midrule
3 & 2  & {\em yes} & 53 & 8 & 129,344.90 &165,353.30 \\
3 & 2 & {\em no} & 49 & 10 & 123,023.90 & 165,573.40 \\
\\
10& 10  &{\em yes} &   50 & 9 & 124,815.80 & 165,320.90\\
10 & 10 & {\em no} & 53 & 8 & 129,388.40 & 165,396.90\\
\\
20 &10 & {\em yes} & 50 & 9 & 124,816.50 &165,321.50\\
20 &10 & {\em no}  & 50 & 9 & 124,835.10 & 165,340.10\\
\\
\bf 30 &\bf  9  & {\em yes} & \bf 50 & \bf 9 &\bf 124,799.10 &\bf 165,304.20\\ 
30 &9 & {\em no} & 51 & 9 & 127,332.90 & 165,486.10\\
\\
  75 &  8 & {\em yes} & 53 & 8 & 129,351.60&165,360.00 \\
  75 & 8 & {\em no } & 53 & 8 & 129,365.90 & 165,374.40\\
 \bottomrule
\end{tabular}
\caption{\em Comparison between the results of the inclusion-exclusion algorithm
for item selection with respect to different sizes of the initial set of clustering items (in boldface are the quantities corresponding to the best solution in terms of $BIC_{tot,k}(\hat{\cg A})$) }\label{tab:DeR}}\vspace*{0.25cm}
\end{table}

\subsection{Validation by resampling}

Given the nature of the search algorithm and the complexity of the study we are dealing with,
the selected number of latent classes and the final set of clustering items may also be
sensitive to the specific data used for the analysis. In order to address this issue, we 
validate the results obtained above by sampling with replacement, from the original
dataset,  $B=99$ random samples of the same size $n$ of the original one.
 
For each sample, we then select the optimal number of latent classes, denoted by $k_b$,
corresponding to the minimum $BIC_{k_b}(\cg I)$, $k_b = 2,\ldots, k_{\max}$, $b=1,\ldots,B$. As done for the original
data, we order the full set of items on the basis of the variability of their estimated
conditional response probabilities across classes with the aim of selecting the 
initial subset of items, $\cg A^{(0)}_b$, and we apply the item selection algorithm with random check. For each sample, we consider a size of the initial set equal to 30 items,
which give the best results, in terms of $BIC_{tot,k}$, in the application to the original data.

In Table \ref{tab:boot} we summarize the results of the above validation procedure. In particular, 
for each item in the full set, we report the number of times that it has been 
selected with respect to the different starting set in the original data, if it has been included 
in the best solution (obtained with a starting set of 30 items as illustrated in Table \ref{tab:DeR}), and
the number of times that the item has been selected with respect to the different random samples obtained in the 
validation procedure. From the table, we observe that 45 items are always included in the different final
solutions, both with respect to different specifications of the initial set of items and with respect to the different random samples 
in the validation procedure. Moreover, items 14 and 68 are always included apart from one and three random samples. 
On the other hand, we note that 7 items are never included in the final solutions, whereas 7 items are included only 
in very few solutions provided by the random samples (see items 19, 22, 28, 36, 39, 72, and 73) .  The remaining 
14 items are in intermediated situations. In conclusion, more than three-quarter of the random samples
confirm the results obtained by the item selection algorithm.

\begin{table}[ht!]\centering\vspace*{0.25cm}
\small{
\begin{tabular}{rrrrr|rrrrr}
\toprule
section&  $j$ & \#sel. & best & \#resamp &section&  $j$ & \#sel. & best & \#resamp  \\
   \midrule
CC	&	1	&	5	&	$\times$ 	&	99	&	HBD	&	39	&		&		&	4	\\
CC	&	2	&	5	&	$\times$	&	99	&	ADL	&	40	&	5	&	$\times$	&	99	\\
CC	&	3	&	5	&	$\times$	&	99	&	ADL	&	41	&	5	&	$\times$	&	99	\\
CC	&	4	&	5	&	$\times$	&	99	&	ADL	&	42	&	5	&	$\times$	&	99	\\
CC	&	5	&	5	&	$\times$	&	99	&	ADL	&	43	&	5	&	$\times$	&	99	\\
CC	&	6	&	5	&	$\times$	&	99	&	ADL	&	44	&	5	&	$\times$	&	99	\\
CC	&	7	&	5	&	$\times$	&	99	&	ADL	&	45	&	5	&	$\times$	&	99	\\
CC	&	8	&	5	&	$\times$	&	99	&	ADL	&	46	&	5	&	$\times$	&	99	\\
CC	&	9	&	5	&	$\times$	&	99	&	ADL	&	47	&	5	&	$\times$	&	99	\\
CC	&	10	&	5	&	$\times$	&	99	&	ADL	&	48	&	5	&	$\times$	&	99	\\
CC	&	11	&	5	&	$\times$	&	99	&	ADL	&	49	&	5	&	$\times$	&	99	\\
CC	&	12	&	5	&	$\times$	&	99	&	ADL	&	50	&	5	&	$\times$	&	99	\\
CC	&	13	&	5	&	$\times$	&	99	&	ADL	&	51	&	5	&	$\times$	&	99	\\
AVF	&	14	&	5	&	$\times$	&	98	&	ADL	&	52	&	5	&	$\times$	&	99	\\
AVF	&	15	&	5	&	$\times$	&	99	&	ADL	&	53	&	5	&	$\times$	&	99	\\
AVF	&	16	&	5	&	$\times$	&	99	&	ADL	&	54	&	5	&	$\times$	&	99	\\
AVF	&	17	&	5	&	$\times$	&	99	&	ADL	&	55	&	5	&	$\times$	&	99	\\
AVF	&	18	&	5	&	$\times$	&	99	&	ADL	&	56	&	5	&	$\times$	&	99	\\
HBD	&	19	&		&		&	5	&	ADL	&	57	&	5	&	$\times$	&	99	\\
HBD	&	20	&	5	&	$\times$	&	96	&	I	&	58	&	5	&	$\times$	&	99	\\
HBD	&	21	&		&		&	12	&	I	&	59	&	5	&	$\times$	&	99	\\
HBD	&	22	&		&		&	2	&	I	&	60	&	5	&	$\times$	&	79	\\
HBD	&	23	&		&		&	10	&	NF	&	61	&	5	&	$\times$	&	99	\\
HBD	&	24	&		&		&	25	&	NF	&	62	&	5	&	$\times$	&	99	\\
HBD	&	25	&		&		&		&	NF	&	63	&		&		&		\\
HBD	&	26	&	2	&		&	59	&	NF	&	64	&	2	&		&	42	\\
HBD	&	27	&		&		&	31	&	NF	&	65	&		&		&		\\
HBD	&	28	&		&		&	1	&	NF	&	66	&	2	&		&	30	\\
HBD	&	29	&		&		&	20	&	DD	&	67	&		&		&	36	\\
HBD	&	30	&		&		&	23	&	DD	&	68	&	5	&	$\times$	&	96	\\
HBD	&	31	&		&		&		&	DD	&	69	&	5	&	$\times$	&	75	\\
HBD	&	32	&	5	&	$\times$	&	99	&	DD	&	70	&		&		&		\\
HBD	&	33	&	5	&	$\times$	&	99	&	DD	&	71	&		&		&		\\
HBD	&	34	&	5	&	$\times$	&	99	&	DD	&	72	&		&		&	9	\\
HBD	&	35	&	5	&	$\times$	&	99	&	SC	&	73	&		&		&	3	\\
HBD	&	36	&		&		&	1	&	SC	&	74	&		&		&		\\
HBD	&	37	&		&		&	17	&	SC	&	75	&	5	&	$\times$	&	99	\\
HBD	&	38	&	5	&	$\times$	&	99	&										\\					
\bottomrule
\end{tabular}
\caption{\em Final results of the item selection algorithm (with random check) and of the validation by resampling. 
$j$ is the item index, \#sel. is the number of times that item $j$ has been selected with respect to the different starting sets, best indicates if item included in the best solution, \#resamp. is the number of times that item $j$ has been selected with respect to the different samples.}\label{tab:boot}}\vspace*{0.25cm}
\end{table}

With respect to the sections of the questionnaire, we observe that
all items referred to sections CC, AVF, and ADL and two items of section I are retained in
$\hat{\cg A}$. On the contrary, most of the excluded items belongs to sections DD, HBD, NF and SC.

\subsection{Parameter estimates}\label{sec:estimates}

With the aim of evaluating the performance of the nursing homes, the adopted approach also allows us to cluster their patients into the different latent classes according to their health conditions on the basis of the parameter estimates. 
This may be useful to describe the case-mix of the nursing homes
and to estimate their ability of retaining patients in the groups corresponding to better health conditions.

In this section, we report the estimation results based on the best solution, in terms of $BIC_{tot,k}(\hat{\cg A})$,
provided by the inclusion-exclusion algorithm, which selects 50 items with $\hat{k}=9$.
Since the items are categorical, with a different number of
categories, we report the estimated conditional response
probabilities, $\hat{\lambda}_{j|u}(y)$, by assigning an
equally-spaced score between 0 and 1 to the different response
categories. Then, we compute the average of the scores,
weighted with the corresponding response probabilities. This amounts
to computing the following {\em item mean score}
\[ 
\hat{\mu}_{j|u} = \frac{1}{l_j-1}\sum_y
(y-1)\hat{\lambda}_{j|u}(y),\quad j \in \hat{\cg A},\; u=1,\ldots,\hat{k},\; y=0,\ldots,l_j-1.
\]
In particular, a value of $\hat{\mu}_{j|u}$ close to 0 corresponds
to a low probability of suffering from a certain pathology, whereas
a value close to 1 corresponds to a high probability of suffering
from the same pathology. To summarize these results, we also compute the  {\em section mean score}
$\hat{\bar{\mu}}_{d|u}$ as the average of $\hat{\mu}_{j|u}$ for the
items in $\hat{\cg A}$ composing each section $d$ of the
questionnaire.

In order to have a clearer interpretation of the results,
we order the latent classes on the basis of the values of
$\hat{\bar{\mu}}_{d|u}$ assumed in the section denoted by ADL
(Activity of Daily Living) of the questionnaire. This is the section with the
highest difference between the maximum and the minimum value of the
section mean score $\hat{\bar{\mu}}_{d|u}$ across classes.

For each latent class, Table~\ref{tab:average} shows the values of
$\hat{\bar{\mu}}_{d|u}$ and the estimated class weights
$\hat{\pi}_u$, together with the difference between the maximum and
the minimum value of $\hat{\bar{\mu}}_{d|u}$ for each section of the
questionnaire. As we can note, the latter is high for sections ADL, CC, I, and AVF, and low for the remaining sections. The smallest among
these differences is observed for section DD, which, consequently,
tends to discriminate less between subjects with respect to the
other sections. The first latent
class, that includes around 17\% of subjects, corresponds to the best health
conditions with respect to all the pathologies measured by the
sections of the questionnaire, apart from section NF, DD and SC. On the other hand, the 9th latent class, which
includes about 11\% of patients, corresponds to cases with the worst
health conditions for almost all the pathologies. Intermediate
classes show a different case-mix depending on the section mean score
pattern. For instance, in the 3rd class are included patients with severe cognitive conditions (CC) and
consistent impairment referred to sections AVF, HBD, and I. Moreover, in the same class we also register 
a moderate impairment of the activities of daily living (ADL).
In the 7th class are instead included patients with worse conditions than those assigned to the 3rd class; in particular, 
in addition to section CC, we also register a severe impairment of the incontinence conditions (section I),
and a worsening of the pathologies measured by sections ADL, NF and SC. 

\begin{table}[h!]\centering\vspace*{0.25cm}
\small{
\begin{tabular}{@{}rrrrrrrrrc@{}}
\toprule & \multicolumn7{c}{$d$}\\
 \cmidrule{2-9} & \multicolumn1c{1}
& \multicolumn1c{2} & \multicolumn1c{3} &
\multicolumn1c{4} & \multicolumn1c{5}& \multicolumn1c{6} & \multicolumn1{c}{7}& \multicolumn1{c}{8}&    \\
\multicolumn1c{$u$} & \multicolumn1c{(CC)}
& \multicolumn1c{(AVF)} & \multicolumn1c{(HBD)} &
\multicolumn1c{(ADL)} & \multicolumn1c{(I)}& \multicolumn1c{(NF)}& \multicolumn1{c}{(DD)} & \multicolumn1{c}{(SC)}&  $\hat{\pi}_u$  \\
\midrule
\multicolumn1c{1}	&	0.041	&	0.098	&	0.066	&	0.090	&	0.230	&	0.085	&	0.392	&	0.047	&	0.169	\\
\multicolumn1c{2}	&	0.361	&	0.231	&	0.205	&	0.131	&	0.347	&	0.078	&	0.390	&	0.031	&	0.105	\\
\multicolumn1c{3}	&	0.694	&	0.444	&	0.464	&	0.238	&	0.688	&	0.168	&	0.398	&	0.056	&	0.084	\\
\multicolumn1c{4}	&	0.143	&	0.178	&	0.089	&	0.315	&	0.407	&	0.148	&	0.377	&	0.100	&	0.096	\\
\multicolumn1c{5}	&	0.596	&	0.360	&	0.292	&	0.527	&	0.776	&	0.215	&	0.338	&	0.067	&	0.098	\\
\multicolumn1c{6}	&	0.079	&	0.141	&	0.078	&	0.628	&	0.609	&	0.125	&	0.406	&	0.145	&	0.103	\\
\multicolumn1c{7}	&	0.757	&	0.593	&	0.357	&	0.680	&	0.894	&	0.337	&	0.371	&	0.172	&	0.102	\\
\multicolumn1c{8}	&	0.479	&	0.328	&	0.180	&	0.739	&	0.844	&	0.240	&	0.401	&	0.200	&	0.129	\\
\multicolumn1c{9}	&	0.735	&	0.666	&	0.259	&	0.895	&	0.888	&	0.439	&	0.380	&	0.315	&	0.114	\\
\midrule
\footnotesize{$\max_{u}(\hat{\bar{\mu}}_{d|u}) - \min_u(\hat{\bar{\mu}}_{d|u})$}     	&	0.716	&	0.568	&	0.398	&	0.805	&	0.664	&	0.360	&	0.068	&	0.284	&		\\
 \bottomrule

\end{tabular}
\caption{\em Estimated section mean score, $\hat{\bar{\mu}}_{d|u}$,
for each latent class $u$ and each section $d$ of the questionnaire,
together with the estimated weights $\hat{\pi}_u$ and the difference
between the largest and the smallest estimated section mean score
for each section, under the latent ignorability
assumption.}\label{tab:average}}
\end{table}

\section{Conclusions}\label{sec:concl}

In this paper, we illustrate the application of an algorithm for item selection, when items are used for 
clustering purposes, which is based on the latent class (LC) model
\citep{laza:50,laza:hen:68,good:74}. This algorithm 
closely follows the one 
proposed by \cite{dean:raftery:2010}, and aims at finding the optimal subset of
items useful for clustering searching for the best result in terms of the Bayesian Information Criterion \citep[BIC,][]{schw:78}. 

More in detail, we illustrate an application based on a
dataset collected within the Italian project named ULISSE
\citep{lattanzio2010}, regarding the quality-of-life of elderly
hosted in nursing homes.
As typically happens in such a context, the questionnaire used to collect the data is made of a large
number of polytomous items. This may lead to a lengthy and
expensive administration of the questionnaire and may induce the
respondents to provide inaccurate responses. In this respect, the algorithm for item
selection we illustrate may lead to a sensible reduction of the
number of items for clustering purposes. Moreover, by removing irrelevant or noise items,
it may improve the performance of the model-based clustering procedure and
the accuracy of the choice of the number of latent classes.
The adopted algorithm extends the inclusion-exclusion algorithm proposed by \cite{dean:raftery:2010},
by including an additional step, which we call random check, aimed at initializing, with a large number
of random starting values, the estimation algorithm, so as to prevent the problem
of the multimodality of the likelihood.

In the present application to the ULISSE dataset, we also perform a sensitivity
analysis of the final solution with respect to different specifications of the initial set 
of clustering items and a validation of the results by means of a resampling procedure.
The results confirm that the random check allows us to increase the chance of reaching the
global maximum of the log-likelihood, especially in the presence of complex models 
characterized by a large number of items and estimated latent classes. 
Moreover, the validation procedure may be useful in applications concerning complex phenomena, 
where the results may be sensitive to the specific data used in the analysis and may be 
affected by potential outliers in the respondents.
 
The best result, in terms of BIC, leads to
selecting 50 items, out of the 75 considered, and 9
latent classes. This reduction implies clear advantages in terms of setting up a
questionnaire which may be more easily administered, especially in a longitudinal context in which we have
repeated measurements.
Most of the selected items belong to sections of the questionnaire referred
to cognitive conditions, auditory and view fields, activities of daily living and incontinence.    
The remaining sections are the ones that tend to discriminate less between subjects in the estimated latent classes.

Once the optimal subset of items has been selected together with the corresponding number of latent
classes, the estimation results may be used to assign subjects to
homogenous classes, which is one of the main aim of the LC model. 
This may have important implications in the context of the ULISSE project, 
where patients are assigned to
different latent classes corresponding to different levels of impairment. This is useful for
evaluating the long-term nursing homes performance with respect to
their ability in improving the patientsÕ health conditions or
in delaying their worsening.

\begin{table}[th!]\begin{flushleft}{\bf \large{Appendix}}\end{flushleft}
\centering
 \tiny{
\begin{tabular}{ccl}
\toprule
\bf $j$ & \bf $\#$ cat.  & \bf item description                           \\
\midrule
\multicolumn3c{Section CC} \\
\midrule
01    &  2  &  Short-term memory   (0 = ``recalls what recently happened (5 minutes)'', 1 = ``does not recall'')\\
02    &  2  &  Long-term memory (0 = ``keeps some past memories green'', 1 = ``does not keep some past memories green'') \\
03    &  2  &  Memory status (0 = ``recalls the actual season'', 1 = ``does not recall the actual season'')                   \\
04    &  2  &  Memory status (0 = ``recalls where is his room'', 1 = ``does not recall where is his room'')                   \\
05    &  2  &  Memory status  (0 = ``recalls the names and faces of the staff'', 1 = ``does not recall the names and faces of the staff'')  \\
06   &  2  &  Memory status (0 = ``recalls where he is'',  1 =``does not recall where he is'')                     \\
07   &  4  &  Decision about his daily activities (from 0 = ``independent decisions'' to 3 = ``unable to decide'') \\
08   &  3  &  Easily sidetracked  (from 0 = ``problems absent'' to 2 = ``problems worsened in the last week'')     \\
09   &  3  &  Altered perception or awareness of surrounding (from 0 = ``problems absent'' to 2 = ``problems worsened in the last week'') \\
10   &  3  &  Disorganized speech (from 0 = ``problems absent'' to 2 = ``problems worsened in the last week'')       \\
11   &  3  &  Restlessness movements (from 0 = ``problems absent'' to 2 = ``problems worsened in the last week'')     \\
12   &  3  &  Lethargic spans (from 0 = ``problems absent'' to 2 = ``problems worsened in the last week'')       \\
13  &  3  &  Change in the cognitive conditions during the day (from 0 = ``problems absent'' to 2 = ``problems worsened in the last week'')    \\
\midrule \multicolumn3c{ Section AVF} \\
 \midrule
14 &  4  &  Hearing (from 0 =``no hearing impairment'' to 3 = ``severe hearing impairment'')    \\
15   &  4  &  Ability to make itself understood  (from 0 = ``understood'' to 3 = ``seldom/never understood'')    \\
16   &  3  &  Clear language (from 0 = ``clear language'' to 2 = ``no language'')        \\
17  &  4  &  Ability to understand others (from 0 = ``understands'' to 3 = ``seldom/never understands'')    \\
18  &  5  &  Sight in conditions of adequate lighting (from 0 = ``no sight impairment'' to 4 = ``severe sight impairment'') \\
\midrule
\multicolumn3c{ Section HDB} \\
\midrule
19  &  3  &  Negative statements (from 0 = ``symptom not showed'' to 2 = ``symptom daily showed'')       \\
20  &  3  &   Repetitive questions (from 0 = ``symptom not showed'' to 2 = ``symptom daily showed'')       \\
21    &  3  &   Repetitive verbalizations (from 0 = ``symptom not showed'' to 2 = ``symptom daily showed'')       \\
22   &  3  &   Persistent anger with himself or others (from 0 = ``symptom not showed'' to 2 = ``symptom daily showed'')      \\
23     &  3  &   Self deprecation disesteem (from 0 = ``symptom not showed'' to 2 = ``symptom daily showed'')      \\
24   &  3  &   Fears that are not real (from 0 = ``symptom not showed'' to 2 = ``symptom daily showed'')       \\
25    &  3  &   To believe himself to be dying (from 0 = ``symptom not showed'' to 2 = ``symptom daily showed'')     \\
26     &  3  &   To complain about his health (from 0 = ``symptom not showed'' to 2 = ``symptom daily showed'')     \\
27     &  3  &   Repeated events anxiety (from 0 = ``symptom not showed'' to 2 = ``symptom daily showed'')\\
28    &   3  &   Unpleasant mood in morning (from 0 = ``symptom not showed'' to 2 = ``symptom daily showed'')    \\
29   &   3  &   Insomnia/problems with sleep (from 0 = ``symptom not showed'' to 2 = ``symptom daily showed'')    \\
30 &   3  &   Expressions of sad-faced (from 0 = ``symptom not showed'' to 2 = ``symptom daily showed'')       \\
31  &    3  &   Easily tears  (from 0 = ``symptom not showed'' to 2 = ``symptom daily showed'')      \\
32  &    3  &   Repetitive movements (from 0 = ``symptom not showed'' to 2 = ``symptom daily showed'')     \\
33  &   3  &   Abstention from activities of interest (from 0 = ``symptom not showed'' to 2 = ``symptom daily showed'')       \\
34  &     3  &   Reduced local interactions (from 0 = ``symptom not showed'' to 2 = ``symptom daily showed'')       \\
35  &    4  &   To wander aimlessly (from 0 = ``problem absent'' to 3 = ``problem daily encountered'')      \\
36  &   4  &    Offensive language (from 0 = ``problem absent'' to 3 = ``problem daily encountered'')    \\
37  &     4  &    Physically aggressive (from 0 = ``problem absent'' to 3 = ``problem daily encountered'')       \\
38  &    4  &    Socially inappropriate behavior (from 0 = ``problem absent'' to 3 = ``problem daily encountered'')      \\
39  &      4  &    To refuse assistance (from 0 = ``problem absent'' to 3 = ``problem daily encountered'')     \\
\midrule \multicolumn3c{ Section ADL} \\
\midrule
40  &  5  &   Moving to/from lying position (from 0 = ``independent'' to 4 = ``totally dependent'')    \\
41  &   5  &   Moving to/from bed, chair, wheelchair (from 0 = ``independent'' to 4 = ``totally dependent'')     \\
42  &  5  &   Walking between different points within the room (from 0 = ``independent'' to 4 = ``totally dependent'')     \\
43  &   5  &   Walking in the corridor (from 0 = ``independent'' to 4 = ``totally dependent'')       \\
44  & 5  &   Walking into the nursing home ward (from 0 = ``independent'' to 4 = ``totally dependent'')       \\
45  &  5  &   Walking outside the nursing home ward (from 0 = ``independent'' to 4 = ``totally dependent'')    \\
46  &   5  &   Dressing (from 0 = ``independent'' to 4 = ``totally dependent'')     \\
47  &  5  &   Eating  (from 0 = ``independent'' to 4 = ``totally dependent'')      \\
48  &  5  &   Using the toilet room (from 0 = ``independent'' to 4 = ``totally dependent'')       \\
49  &  5  &   Personal hygiene  (from 0 = ``independent'' to 4 = ``totally dependent'')    \\
50  &   5  &   Taking full-body bath/shower (from 0 = ``independent'' to 4 = ``totally dependent'')     \\
51  &  4  &   Balance problems (from 0 = ``does not have balance problems'' to 3 = ``needs physical assistance'')       \\
52  &   3  &   Mobility in the neck (0 = ``no limitation'', 1 = ``unilateral limitation'', 2 = ``bilateral limitation'')      \\
53  &    3  &   Mobility in the arm including shoulder or elbow (0 = ``no limitation'', 1 = ``unilateral limitation'', 2 = ``bilateral limitation'')        \\
54  &   3  &   Movements of the hand including wrist or finger     (0 = ``no limitation'', 1 = ``unilateral limitation'', 2 = ``bilateral limitation'')    \\
55  &   3  &   Mobility in the leg and hip (0 = ``no limitation'', 1 = ``unilateral limitation'', 2 = ``bilateral limitation'')         \\
56  &      3  &   Mobility in the foot and ankle (0 = ``no limitation'', 1 = ``unilateral limitation'', 2 = ``bilateral limitation'')         \\
57  &     3  &   Other movements (0 = ``no limitation'', 1 = ``unilateral limitation'', 2 = ``bilateral limitation'')          \\
\midrule \multicolumn3c{ Section I}\\
\midrule
58   &  5  &     Fecal incontinence (from 0 = ``continence'' to 4 = ``incontinence'')      \\
59       &  5  &     Urinary incontinence (from 0 = ``continence'' to 4 = ``incontinence'')        \\
60     &  2  &     Elimination of feces (0 = ``adequate'', 1 = ``not adequate'')      \\
\midrule \multicolumn3c{ Section NF}\\
\midrule
61  &   2  &      Chewing problem (0 = ``no problem'', 1 = ``problems'')        \\
62  &  2  &      Swallowing problem  (0 = ``no problem'', 1 = ``problems'')         \\
63  &  2  &      Mouth pain (0 = ``no problem'', 1 = ``problems'')           \\
64  &  2  &      Taste of many foods (0 = ``does not complain'', 1 = ``complains'')       \\
65  &  2  &      Hungry  (0 = ``does not complain'', 1 = ``complains'')   \\
66  & 2  &      Food on his plate (0 = ``does not leave it'', 1 = ``leaves it'')      \\
\midrule \multicolumn3c{ Section DD}\\
\midrule
67  &   2  &      Debris present in mouth prior to going to bed at night (0 = ``problem absent'', 1 = ``problem present'') \\
68  &   2  &      Dentures/removable bridge (0 = ``absent'', 1 = ``present'') \\
69  &  2  &     Some/all natural teeth lost and does not have/does not use dentures (or partial plates) (0 = ``problem absent'', 1 = ``problem present'') \\
70  &   2  &     Broken, loose, or carious teeth  (0 = ``problem absent'' 1 = ``problem present'')         \\
71  &   2  &     Inflamed gums, swollen or bleeding gums, oral abscesses, ulcers or rashes  (0 = ``problem absent'', 1 = ``problem present'') \\
72  &  2  &     Dentures or removable bridge daily cleaned by resident or staff (0 = ``absent'', 1 = ``present'') \\
     \midrule \multicolumn3c{ Section SC}\\
\midrule
73  &   5 &      Pressure ulcer (from 0 = ``no pressure ulcer'' to 4 = ``stage 4'')   \\
74  &   5 &      Stasis ulcers  (from 0 = ``no pressure ulcer'' to 4 = ``stage 4'')    \\
75  &    2 &      Resolved or cured ulcer (0 =``absent'', 1 = ``present'')    \\
\bottomrule
\end{tabular}}\small{
\caption{\em Description of the full set of items.}\label{tab:list}}
\end{table}%
\bibliography{refer_ulisse}
\bibliographystyle{apalike}

\end{document}